# Optical Spin Sorting Chain

*Tatsuki Hinamoto,\*,†, Minoru Fujii†, and Takumi Sannomiya,\*,‡,§*

†Department of Electrical and Electronic Engineering, Graduate School of Engineering, Kobe University, Rokkodai, Nada, Kobe 657-8501, Japan

‡Department of Materials Science and Technology, Tokyo Institute of Technology, 4259 Nagatsuta Midoriku, Yokohama 226-8503, Japan

§JST-PRESTO, 4-1-8 Honcho, Kawaguchi, Saitama 332-0012, Japan

ABSTRACT

Transverse spin angular momentum of light is a key concept in recent nanophotonics to realize unidirectional light transport in waveguides by spin-momentum locking. Herein we theoretically propose subwavelength nanoparticle chain waveguides that efficiently sort optical spins with engineerable spin density distributions. By arranging high-refractive-index nanospheres of different sizes in a zigzag manner, directional optical spin propagation is realized. The origin of the efficient spin transport is revealed by analyzing the dispersion relation and spin angular momentum density distributions. In contrast to conventional waveguides, the proposed asymmetric waveguide can spatially separate up- and down-spins and locate one parity inside and the other outside the structure. Moreover, robustness against bending the waveguide and its application as an optical spin sorter are presented. Compared to previous reports on spatial engineering of local spins in photonic crystal waveguides, we achieved substantial miniaturization of the entire footprint down to subwavelength scale.

Advances in the field of nanophotonics have paved the way toward highly integrated photonic circuits by miniaturizing optical components. Recently, recognition of the spin-momentum locking (SML) of light, in analogy to the quantum spin Hall effect, is revolutionizing the field.[1] SML is the key concept of chiral light-matter interactions, where the direction of momentum is fundamentally locked by the polarization state of light.[2] Using this idea, spin-photon interfaces are now capable of uniquely determining the information transfer direction from or to a quantum emitter according to its spin. There are versatile fields that gain benefits from chiral light-matter interactions, ranging from valleytronics to quantum information processing.[3,4] In particular, quantum network designs would be substantially accelerated through the development of basic device elements that works as qubits, memories, nodes, paths, routers, isolators, and so on.

SML is a universal phenomenon accompanied by evanescent waves;[2] in principle, no complicated structures are required to realize chiral interactions. Indeed, SML has been demonstrated in various photonic structures, such as metal surfaces[5–7], fibers[8], strips[9–11], and photonic crystal waveguides[12,13]. In such symmetric one-dimensional (1D) waveguides, SML is typically achieved by positioning an individual scatterer beneath the waveguides or by embedding or placing a light source (*e.g.,* quantum dots) at a position offset from the propagation so that spins with different parities have different spatial or energetical distribution. This however requires sophisticated and accurate positioning of the emitter since the spin densities are equally distributed in the same material and a slight displacement in the material can flip the SML direction. On the other hand, in the case of photonic crystal waveguides, the local spins in the waveguide itself have been spatially engineered.[12] Although its large degree of freedom to engineer optical properties is useful, the two-dimensional (2D) nature of photonic crystals can be a limitation in highly integrated photonic circuits since the entire footprints spanning over dozens of periods (typically >10$\lambda$) are required to constitute photonic bands.

Herein, we propose a novel 1D waveguide in the form of a chain, which can highly miniaturize the SML waveguides with engineerable spin densities down to a subwavelength scale. The design of the waveguide is based on linearly aligned 1D nanoparticle chains made of high-index dielectrics. In such dielectric chain structures, it is known that light propagates with low loss (~5 dB/mm[14]) by couplings of Mie resonances resulting from the confinement of light within each nanoparticle.[14–22] They are one of the promising waveguides because of various prominent features: bending stability to 90° corner[19], all-optical modulation with a response time of 50 ps[16], and slow light with group velocity down to 0.03 of the speed of light[17]. Our idea in this work is to introduce asymmetry to the 1D chain waveguide by a side chain to realize an SML waveguide that spatially separates up and down local spins predominantly inside and outside waveguides. In contrast to conventional 1D waveguide geometries, the spatial distributions of local spins can be tailored by simple structural parameters, such as the particle size, position, and gap. Furthermore, the discontinuous nature of the chain structure offers a facile connection to various functional photonic components including isolators and routers, which we also demonstrate in this letter.

Figure 1a schematically visualizes the concept of this work. The proposed waveguide consists of silicon nanospheres of different sizes linearly arranged in a zigzag manner and is excited with a circularly polarized dipole to mimic a quantum dot that has circular polarization states. To study the SML in the chain, we simulate its optical properties using the finite-difference time-domain (FDTD) method. The simulation model is shown in Figure 1b. Silicon spheres with radii $R_1$ and $R_2$ are aligned with a fixed gap length $g$. The number of large spheres ($N$) is fixed to 31, if not specified. The circular dipole is composed of linearly polarized two dipoles oriented along $x$- and $y$-axes, respectively. By setting a $\pi/2$-phase difference between these orthogonal dipoles, the source has $\sigma^\pm$ polarizations ($\bm{E} = E_0(\bm{x} \pm i\bm{y})e^{i\omega t}$). At first, the source is situated at the center of a larger sphere at the middle of the chain (the 16th sphere when $N$ = 31). The spin-sorting performance of the chain is evaluated by light transmission at the rightmost ($T_R$) and leftmost ($T_L$) spheres. We define a starter structural parameter set ($R_1$, $R_2$, $g$) = (120 nm, 80 nm, 5 nm), which we call the "basic parameter set". These spheres have the dipolar Mie resonances (*i.e.*, electric and magnetic dipole modes) around red to near-infrared spectral range. Further details of the simulation method are provided in the Supporting Information.

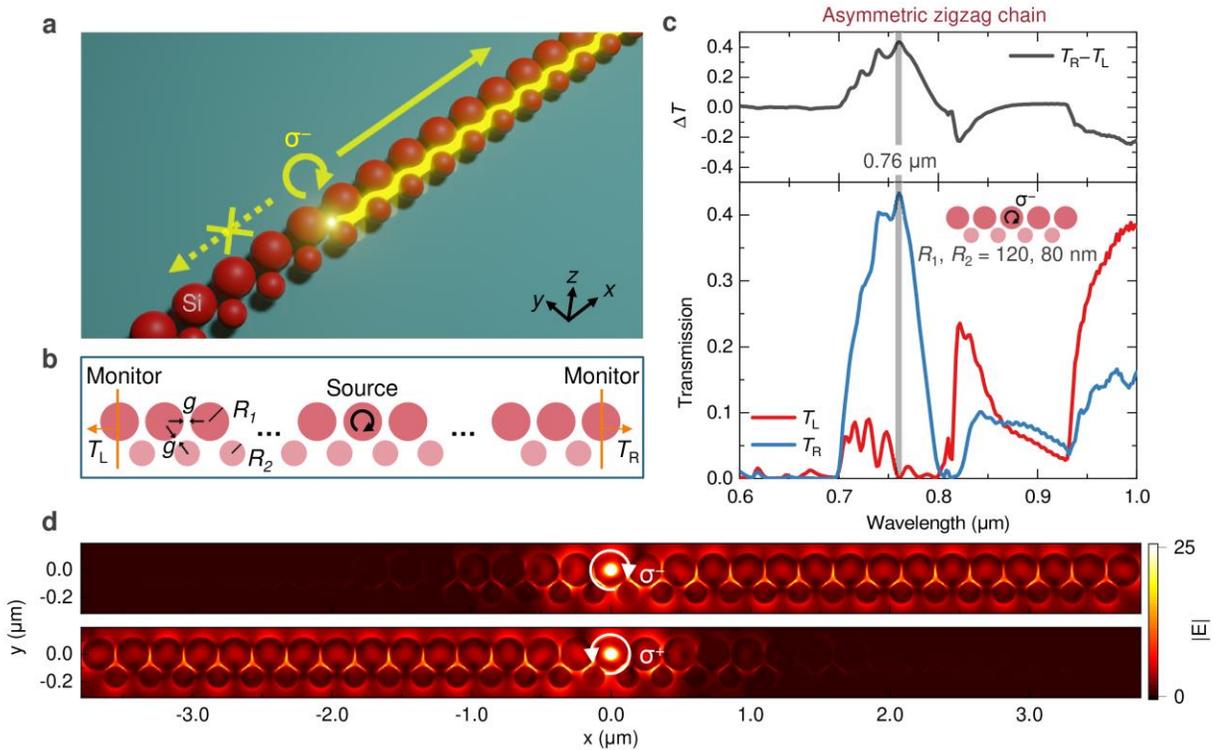

**Figure 1.** Concept of the proposed asymmetric zigzag chain waveguide. (a) Schematic illustration. (b) Simulation configuration for the waveguide consisting of silicon nanospheres. (c) Light

transmission under the excitation with a $\sigma-$ dipole. Transmission spectra in the right ($T_R$; blue curve) and left ($T_L$; red curve) direction are shown on the bottom. The difference $\Delta T = T_R - T_L$ is shown on the top. (d) Electric field distributions at the wavelength of 0.76 μm under $\sigma-$ (top) and $\sigma+$ (bottom) dipole excitation. The calculations in panels c and d are performed with the basic parameter set of $(R_1, R_2, g) = $ (120 nm, 80 nm, 5 nm).

Figure 1c (bottom panel) shows transmission spectra for an asymmetric zigzag chain with the basic structural parameter set under excitation with a $\sigma-$ dipole. The spectra show different spectral responses between the right (blue) and left (red) side propagations. For example, it indicates that ~40% of coupled light propagates to the right in the wavelength range of 0.7 – 0.8 μm (≡ the first band), while less than 10% reach the left side. In the wavelength above 0.93 μm (≡ the second band) opposite behavior is observed, that is, the light preferentially propagates to the left. The difference of the transmission, $\Delta T = T_R - T_L$, is plotted in the top panel, showing the higher selectivity in the first band. The highest contrast is observed at the wavelength of 0.76 μm. Note that the highest $\Delta T$ wavelength slightly differs depending on $N$ because of the formation of a band within a finite number of periods (Figure S1). Even after averaging the transmission over $N$ (*i.e.*, $N$ dependent factors are eliminated), 97% of the transmitted light reaches the right at the highest $\Delta T$ wavelength. Besides, the operating wavelength can be readily tuned by the sphere size because of the scaling nature of the Mie resonance (Figure S2). An electric field profile at $\lambda = 0.76$ μm is shown on the top panel of Figure 1d and clearly depicts unidirectional propagation of light towards the right without significant decay. Because the system does not break time-reversal symmetry, the propagating direction can be switched oppositely under $\sigma^+$ dipole excitation as shown on the bottom panel. These behaviors unambiguously evidence the existence of SML.

To further understand the effect of symmetry breaking by the side chain, we present the side chain size ($R_2$) dependence of the transmission spectrum with a fixed large chain size ($R_1$ = 120 nm) and gap length ($g$ = 5 nm) in Figure 2. Figures 2a and b show two-dimensional (2D) transmission maps when $R_2$ is changed from 0 nm (a linear 1D chain without the side chain) to 120 nm ($R_1 = R_2$, that is, a symmetric zigzag chain). The corresponding 2D map of $\Delta T$ is shown in Figure 2c. We see that in the first band ($\lambda$ = 0.7 to 0.8 μm) the SML arises ($\Delta T \sim +0.5$) at around $R_2 = 80$ nm, corresponding to the radius ratio $R_2/R_1 \sim 0.66$, while the opposite parity SML appears ($\Delta T \sim -0.5$) above $R_2 = 90$ nm in the second band ($\lambda > 0.93$ μm).

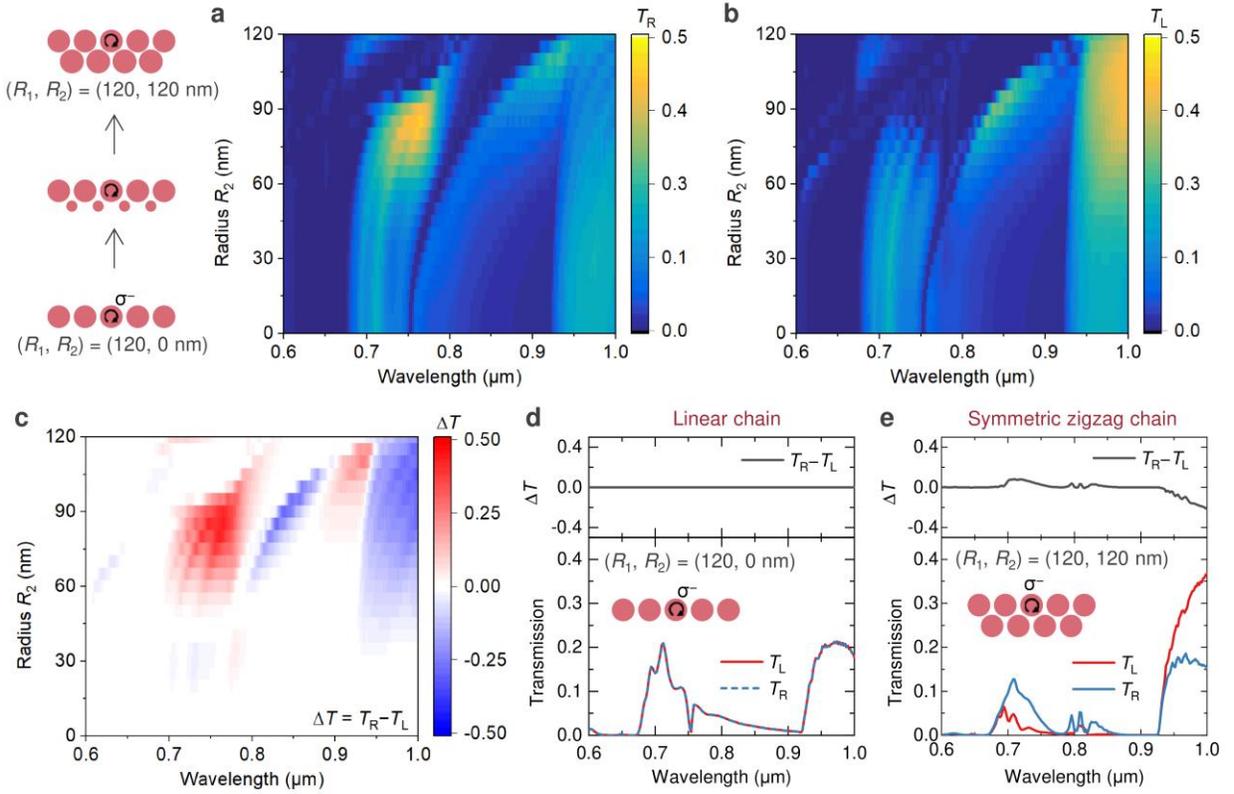

**Figure 2.** Radius ratio dependent light propagation in the asymmetric zigzag chain waveguide for $R_2$ from 0 to 120 nm ($R_1$ and $g$ are fixed to 120 and 5 nm, respectively). (a,b) Transmission on the (a) right and (b) left edges and (c) transmission difference $\Delta T = T_R - T_L$. Transmission spectra of (d) the linear chain and (e) the symmetric zigzag chain. The plots in panels d and e correspond to the profiles in panels a-c at $R_2 = 0$ and 120 nm, respectively. The excitation dipole is a $\sigma-$ dipole excitation in all cases.

For more detailed analysis, the transmission spectra of the symmetric waveguides, *i.e.*, the linear ($R_2 = 0$ nm) and the symmetric zigzag chains ($R_2 = 120$ nm), are extracted and plotted in Figures 2d and e, respectively. In the case of the linear chain, $T_R$ and $T_L$ are identical because of its symmetry. The first and second transmission bands are still observed around 0.7 and 1.0 μm, with a slight blue-shift compared to the basic structure shown in Figure 1. We notice that a sum of the peak values ($T_R + T_L \sim 0.4$) is equivalent to the peak $T_R$ (first band) or $T_L$ (second band) of the asymmetric zigzag chain (Figure 1c). This means that adding the side chain does not degrade the transmission efficiency. In the symmetric zigzag chain (Figure 2e), we see significantly degraded transmission in the first band. In this configuration, the directionality originates from the asymmetrically positioned excitation source.[9] Although the total transmission is comparable to the asymmetric zigzag chain for the second band, the $\Delta T$ ($\sim -0.2$) is much smaller than the asymmetric chain. As another example of structural design, the gap length dependence is provided in the Supporting Information (Figure S3). Besides, we remark that arrays of nanodisks, which can be produced by conventional nanofabrication approaches, have a much more degree of freedom to tailor chracteristics than those of spheres (Figure S4).

Next, we discuss the origin of the SLM behavior in the asymmetric zigzag chain. To this end, we present dispersion relations of the linear and asymmetric zigzag chains in Figure 3. The ordinate and abscissa are a free-space wavenumber ($k_0 = 2\pi / \lambda$) and a Bloch wavenumber ($\beta$), respectively, normalized by a periodicity (*a*) and π. They are simulated for infinite chains using Bloch boundaries in the propagation direction (*x*-axis). Nonrelevant modes in the proposed system, *i.e.*, LM, LM', and TE$_z$' modes, are also included and plotted with orange curves. Mode profiles in panels b and d are calculated at a Bloch wavenumber $\beta a / \pi = 0.8$ and corresponding mode frequencies.

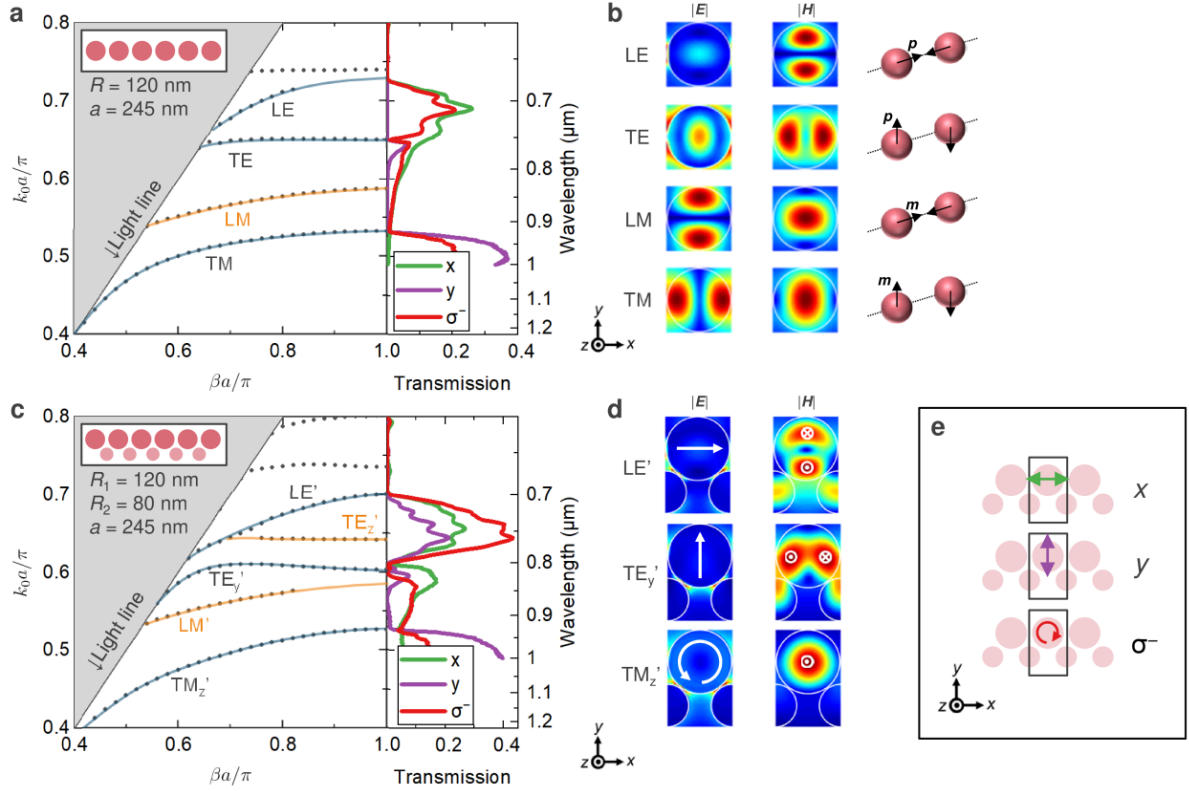

**Figure 3.** Dispersion analysis of the guided modes. (a,c) Dispersion relation of (a) linear chain consisting of large nanoparticle ($R$, $g$ = 120, 5 nm) and (c) Asymmetric zigzag chain with the basic parameter set ($R_1$, $R_2$, $g$ = 120, 80, 5 nm). In both cases, the chains are infinitely long and have the same periodicity of 245 nm ($a = 2R_1 + g$). The shaded gray region shows the energy above the light line. On the right panels, corresponding transmission spectra ($T_R$) are shown. The definition of the excitation dipoles for transmission is shown in panel e. (b,d) Mode profiles at $\beta a / \pi = 0.8$ that correspond to panels a and c (left: electric field; right: magnetic field). The modes shown with orange curves in panels a and c cannot be excited in the geometry of transmission simulations due to a symmetry reason.

Figure 3a shows the dispersion relation of the linear chain. As can be seen, there are several guided modes below the light line. Mode profiles in Figure 3b allow us to assign the modes. In agreement with the previous reports[14,15,18], the first two modes from the low energy side are transverse-magnetic (TM) and longitudinal magnetic (LM) modes, where the magnetic dipole modes are transversely and longitudinally polarized, respectively. The abbreviation follows the convention in this field.[15] The next two bands are transverse-electric (TE) and longitudinal-electric (LE) modes which originate from the electric dipole resonance. By comparing the bands of the dispersion relation with the transmission spectra shown in the side panels under excitation with either *x*- (green) or *y*- (purple) oriented linear or *σ*− circular (red) dipoles, it is seen that the observed first and second transmission bands around 0.7 and 1.0 μm are dominated by LE and TM modes, respectively. LM mode does not appear in the transmission spectrum under the dipole source in the *xy*-plane due to a mismatch of symmetry.

In Figures 3c and d, dispersion relations and mode profiles of the asymmetric zigzag chain are presented. The modes are not purely made of transverse/longitudinal electric/magnetic dipole modes anymore, and thus they are distinguished from the symmetric chain with the primes in the notation. For example, the LE' mode has both LE and TE components due to the interaction with the side chain. For simplicity, we limit the discussion below to the modes that can be excited with a dipole oriented in the *xy*-plane (*i.e.,* bands traced by blue curves). The overall tendency is similar to the linear chain except for a slight low energy shift because of the increased modal volume. The relatively small deviation of the band structure, despite the perturbation of the mode profiles, is consistent with the non-degraded transmission efficiency in the asymmetric zigzag chain as discussed above. The dispersion reveals that the origin of the first transmission bands with SML can be attributed to the LE' mode. Likewise, the second band can be assigned to the TM$_z$' mode.

To obtain more physical insight into the SML in this system, we now discuss the spatial distribution of the transverse optical spin associated with the LE mode. The spin of structured optical fields can be described by spin angular momentum (SAM) densities in the following formalism[23,24]:

$$\boldsymbol{s} = \text{Im}[\varepsilon_0 \mu_r^{-1} \boldsymbol{E}^* \times \boldsymbol{E} + \mu_0 \varepsilon_r^{-1} \boldsymbol{H}^* \times \boldsymbol{H}]/4\omega. \qquad (1)$$

This is the quantity responsible for the generation of SAM. The top panel of Figure 4a shows an electric field profile ($|\boldsymbol{E}|$) of the LE mode ($\lambda = 0.71$ μm) of the linear chain on the *xy*-plane ($z = 0$) under excitation by a linear electric dipole with an orientation along the *x*-axis. Other electromagnetic field components ($E_x$, $E_y$, $H_z$) are presented in the Supporting Information (Figure S5). It is seen that the launched field from the source at the central sphere counter-propagates to both sides. Besides, the electric field is strongly confined around the line of $y = 0$. The normalized SAM density on the same plane is plotted in the bottom panel. Because $E_z$, $H_x$, and $H_y$ are zero at $z = 0$, the SAM density is purely made of electric component and is composed of $s_z$ component in the plane. The red and blue colors in the plot indicate the opposite rotation directions of local fields (*i.e.,* up- and bottom-spins) when the light propagates. In this symmetric linear chain, the local spins are symmetrically distributed on both *y*-positive and negative sides albeit the opposite sign of the spin. In general, we can expect that local sources can efficiently couple to guided modes at the electric field maxima; however, the field distribution indicates the local spin does not exist (*s* = 0) at the electric field maximum at $y = 0$. In other words, ideal spots for spin coupling where both $|\boldsymbol{E}|$ and $|\boldsymbol{s}|$ are both maximized do not exist in the linear chain. Indeed, even if the source is placed at the SAM density maximum, the light transmission is significantly suppressed, leading to $\Delta T < 0.1$ in the LE mode (Figure S6).

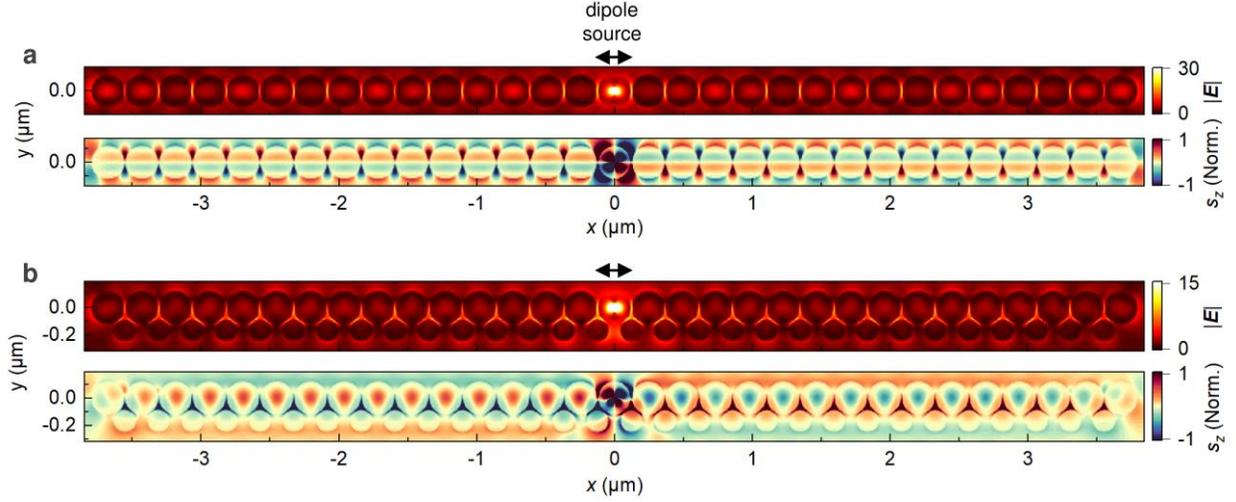

**Figure 4.** Profiles of electric fields (top panels) and normalized spin densities ($s_z$) for (a) the linear chain ($R$, $g$ = 120, 5 nm) and (b) asymmetric zigzag chain ($R_1$, $R_2$, $g$ = 120, 80, 5 nm) under the excitation with a *x*-oriented electric dipole source. The profiles at the wavelengths of the LE and LE' modes (a: $\lambda$ = 0.71 μm; b: $\lambda$ = 0.76 μm) are shown.

The electric field profile of the LE' mode ($\lambda$ = 0.76 μm) of the asymmetric zigzag chain in the same geometry is shown in the top panel of Figure 4b, which resembles the field profile under the circular dipole excitation (Figure 1d). In stark contrast to the linear chain, the SAM density of the asymmetric zigzag chain in the bottom panel is not symmetrically distributed with respect to the *xz*-plane. On the right side of the chain (*i.e.,* for light propagating to the right), down-spins dominate the large sphere, while the gap region is dominated by up-spins. This elucidates the observed directional SAM transport in Figure 1, that is, the light from the $\sigma^-$ dipole positioned in the large sphere selectively propagates in the right direction. Moreover, the electric field maximum nicely overlaps the SAM density maximum at the center of the sphere. Thus, adding the side chain enables us to realize a spatial overlap between the high SAM density and field intensity (electromagnetic density of states), resulting in the efficient chiral coupling of the rotating electric dipole and the propagating mode.

In addition, the asymmetric zigzag chain offers accessible spots of efficient chiral light-matter couplings in the gap region between large and small spheres with highly enhanced electric fields as well as the SAM density. As shown in Figure S7, we see equivalently high transmission and directionality under the excitation at the gap, while the transport direction is opposite. This provides an efficient and easily accessible coupling site to place light sources outside the structure, circumventing technical difficulties to embed quantum dots inside dielectric materials.[9,11,12,25]

Finally, we demonstrate the functionalities of the proposed chain. Figure 5a presents the spin transport over a bent chain with a bending radius of 2.5 μm. The directional transport in the anti-clockwise direction is clearly observed, though the electric field decays as it propagates compared to the straight chain. In contrast to photonic crystal waveguides, where the guiding direction is fixed to the crystal orientation, the truly 1D waveguide allows flexible bending in a continuous manner. In the Supporting Information Figure S8, we provide the light guiding over a sharp corner (~ 60°) in the first and second bands. The structural optimization would be the subject of future work, including the stability to 90° bending such as linear chains.[19]

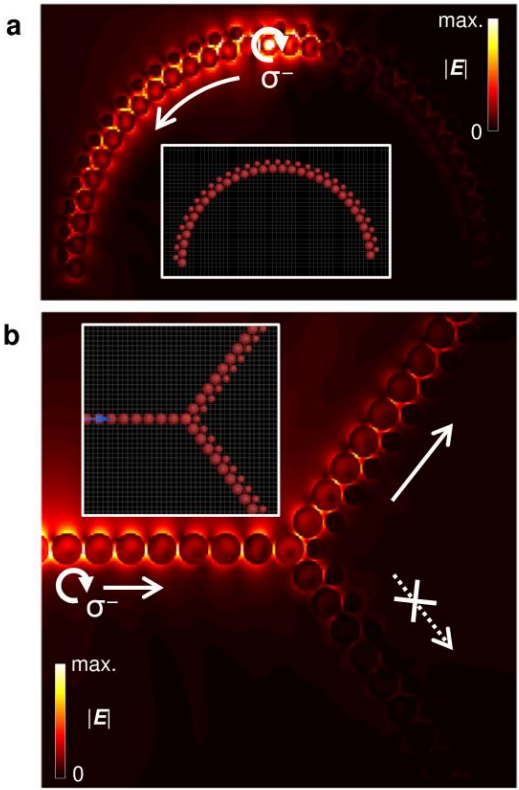

**Figure 5.** Demonstration of functionalities of the asymmetric zigzag chain waveguide ($R_1$, $R_2$, $g$ = 120, 80, 5 nm). (a) Directional light guiding along a bent chain with a bending radius of 2.5 μm ($\lambda$ = 0.764 μm). (b) Chiral light sorter, where the impinging light is sorted into upper or lower branches depending on the circularity of the light ($\lambda$ = 0.783 μm). The excitation source is a $\sigma^-$ dipole.

In addition to the SML waveguiding, the introduction of asymmetry offers new functional elements that cannot be recognized in symmetric systems, for instance, an optical spin sorter (one-way splitter). As shown in Figure 5b, we design a star connection spin-sorter consisting of a linear chain as an input line and two lines of asymmetric zigzag chains as output ports. To generate the selectivity in light propagation depending on the transverse optical spins of impinging light, the asymmetric chains are arranged oppositely for upper and lower branches. The light source is a circular dipole in a linear chain, and $\sigma^-$ polarized light is injected into the junction. Due to the asymmetric distribution of the SAM density in the asymmetric zigzag chain, light is sorted into the upper branch. By switching the rotation of the source, the sorter branch is flipped to the bottom one. Similar junctions that selectively transport circular light into a certain direction have been reported in photonic crystal systems using an approach based on the topology.[26,27] The asymmetric zigzag chains miniaturize such devices down to a sub-wavelength scale, offering great compatibility with integrated photonics. In addition to this routing element, an optical spin isolator, that attenuates injected light of a certain parity (either up or down), can be realized by placing a damping element at one end.

In conclusion, we proposed an asymmetric zigzag chain made of silicon nanospheres as a novel platform for chiral light-matter interactions. Unlike previously studied waveguides whose symmetry is broken by the position of a light source or by an additional coupler, the chain inherently possesses SAM of light. Its large degree of freedom to design structure enables engineering of SAM densities for efficient couplings of local light sources at the electric field maxima. Furthermore, we proposed a novel design concept of spin device elements, such as an optical spin sorter (and router), which offers significant miniaturization of device footprints compared to previously reported similar systems based on photonic crystal waveguides (from > $10\lambda$ to ~ $0.5\lambda$). To the best of our knowledge, this is the first report showing the SML phenomenon in chain systems. The proposed chain may be fabricated by assembling colloidal solutions of silicon nanospheres.[28–31] The presented strategy works as well for nanodisk arrays which are readily producible using standard nanofabrication technologies. Moreover, we expect the current perspective on Mie-tronics may be widely extended by shedding light on SML phenomena associated with Mie resonances.[32–35]

ASSOCIATED CONTENT

**Supporting Information**. The following files are available free of charge.

Methods; number of particles dependence; size dependence with a fixed $R_1$-to-$R_2$ ratio; gap length dependence; spin transport in asymmetric zigzag chains made of nanodisks; detailed mode profiles and spin density distributions; transmission spectra of the linear chain under excitation with a circular dipole positioned with an offset from the center of the structure; transmission spectra of the asymmetric chain for a dipole source positioned in the gap; and demonstration of robustness against sharp corners (PDF)

AUTHOR INFORMATION

**Corresponding Author**

*E-mail: tatsuki.hinamoto@gmail.com

*E-mail: sannomiya.t.aa@m.titech.ac.jp

ACKNOWLEDGMENT

T. H. acknowledges the support under Grant-in-Aid for JSPS Research Fellows. This work was partly supported by JSPS KAKENHI Grants 18J20276, JST PRESTO (JPMJPR17P8), and Murata Science Foundation.

TOC Graphic

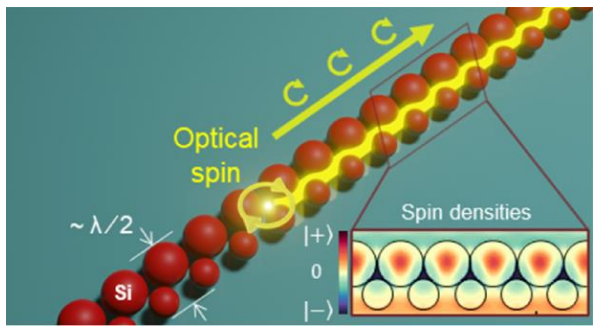